 \documentclass[final,5p,times,twocolumn,authoryear]{elsarticle}

\usepackage{amssymb}
\usepackage{amsmath}
\usepackage{graphicx}
\usepackage{caption}
\usepackage{wrapfig}
\usepackage{multirow}
\usepackage{hyperref}
\usepackage{enumitem}
\usepackage{dblfloatfix}
\usepackage{placeins}
\usepackage{comment}
\usepackage{makecell}
\usepackage[tableposition=top]{caption}
\usepackage{lipsum}
\hypersetup{
  colorlinks  = true, 
  urlcolor    = cyan, 
  linkcolor   = cyan, 
  citecolor   = red 
}

\usepackage{setspace}

\journal{the Journal of Air Transport Management}

\begin{document}

\begin{frontmatter}


\title{Analysing air transport connectivity in Africa through airline decisions: a choice modelling approach}

\author[first]{Khevna Rajput}
\author[first]{Mark Zuidgeest}

\author[first,second]{Philipp Fr\"{o}hlich}
\author[first,third]{Stephane Hess}

\affiliation[first]{organization={Centre for Transport Studies, University of Cape Town},
            addressline={Rondebosch}, 
            city={Cape Town},
            country={South Africa}}

\affiliation[second]{organization={TransSol GmbH},
                       city={Wollerau},country={Switzerland}}
\affiliation[third]{organization={Choice Modelling Centre \& Institute for Transport Studies (ITS), University of Leeds},
            city={Leeds},
            country={United Kingdom}}
            
\begin{abstract}
Transportation models, including air travel models, provide insights into the dynamics of transport supply and demand, which enhances route planning, among others. Against a backdrop of changes in urbanisation, the economy, and geopolitics, decisions by authorities, airlines, and airline consortia lead to changes in the supply of air travel between cities and between countries. Advanced airline choice models create the opportunity to better understand airline preferences, such as route planning, flight frequencies and seat capacity.  Africa is largely overlooked in global air transport research despite rapid urbanisation and rising air traffic across the continent. This growing demand highlights the need to investigate airline behaviour within the continent.  Understanding these preferences not only helps to improve intra-African connectivity but also has the potential to enhance passenger satisfaction by aligning airline supply services more closely with actual travel demand. 
This research tries to understand how seat capacity on routes connecting 267 African cities is distributed across airlines by analysing supply decisions between 2016 and 2022. The paper compiles an Africa-specific database on seat capacity by origin-destination pair and year, using revealed preference data for one week in November each year extracted from the Official Aviation Guide (OAG). Supplemented with socio-economic and historical data such as GDP per capita, market size, predominant language, and colonial ties, the research applies the extended Multiple Discrete-Continuous (eMDC) choice model. 
The model results show that African carriers are, indeed, the ones offering the best connectivity for intra-African travel, but they also show less flexibility in their route networks. As expected, there is significant preferential treatment for national carriers regarding serving their home countries. The study highlights the structural and regulatory barriers that continue to constrain intra-African air connectivity, despite policies such as the Yamoussoukro Decision, designed to liberalise the sector. The dominance of airlines in serving only their home markets suggests a lack of cross-border cooperation and competition, limiting the potential for an integrated, efficient African aviation market. This not only restricts mobility across the continent but also hampers broader economic growth. The model estimates are statistically significant at a 95\% confidence level, supporting the use of these findings for informed policy and strategic decision-making. 
This paper represents the first application of discrete-continuous modelling using capacity-supply data within the African aviation context and offers policy-relevant insights into improving air connectivity. 

\end{abstract}



\begin{keyword}
African aviation \sep airline industry \sep air transport \sep connectivity\sep origin-destination \sep transport modelling \sep choice modelling



\end{keyword}

\end{frontmatter}




\section{Introduction}
\label{introduction}

The airline industry plays a crucial role in fostering economic growth and connecting cities by facilitating trade and mobility. However, in Africa, air transport infrastructure remains underdeveloped in comparison to other continents. Understanding how airline infrastructure and seat capacity are allocated across African origin-destination (OD) pairs is key to improving connectivity and guiding policy interventions. This paper examines air transport connectivity in Africa by analysing capacity-supply data from the Official Aviation Guide (OAG), offering insights into the determinants of airline decisions and the dynamics shaping supply across the continent.

The focus on airline capacity-supply decisions brings a unique side to the study, offering a commercial and operational perspective on air connectivity. Most of the existing research \citep{Hess2005, Kolker2024, Ragab2024} focuses on passenger decisions, whereas this study shifts the focus to understanding how capacity on a given route is allocated across airlines based on economic, historical and geopolitical factors. 

Unlike traditional regression and network demand studies \citep{Burghouwt2013, Bordeaux2024, Kolker2025}, this paper applies advanced discrete-continuous choice modelling to quantify capacity-supply decisions across airlines, taking into account a range of factors including GDP, market size, geographic proximity, colonial ties, language, and external shocks such as COVID-19, amongst others. Contrary to ordinary regression, choice models allow for a more detailed understanding of how different factors shape decision-making, which includes the quantification of utilities associated with different origin-destination attributes, thus offering a more thorough understanding of supply behaviour. By distinguishing between capacities supplied by African and non-African airlines, the paper also explores the extent of regional self-reliance versus external dependency. The results aim to support data-driven policy recommendations for improving air transport integration, network planning, and infrastructure development.

The study uses revealed preference (RP) data, aggregated from actual observed airline supply choices for one week in November across multiple years in the OAG. Such RP datasets are often used in air transport management studies on passenger choices \citep[see e.g.][]{Hess2005}. They can then also be used to develop models to forecast responses, like in a study that looks at impacts of infrastructure upgrades or network extensions resulting from reduced airport access time \citep{Ryley2019}. An important reason for the lack of supply-side work is the absence of data on airline decision-making - the use of the OAG data acts as an important source that we exploit for this purpose.

Ultimately, this paper investigates the extent to which African carriers dominate international intra-African routes. It explores and quantifies the key factors that influence airline supply decisions, particularly their impact on seat allocation decisions across routes using discrete-continuous choice models. Based on that, we discuss how these findings can inform strategic and policy decisions to better connect African countries.

The remainder of this paper is structured as follows. Section \ref{background} provides some background on air transport in Africa, followed by an overview of the literature in Section \ref{literature} on capacity-supply decisions, the Yamoussoukro Decision, and various factors affecting air connectivity. Section \ref{data} provides insight into the data, and Section \ref{method} explains the modelling methodology in detail. The eMDC choice modelling results are presented and interpreted in Section \ref{results}, and the paper concludes with a discussion, summary of key findings and limitations in Section \ref{conclusion}.

\section{Background}\label{background}

Air transport plays an important role in the global economy by facilitating passenger mobility across borders. For African countries, many of which are landlocked, air connectivity offers critical access to global networks. However, the continent continues to face substantial challenges towards a fully integrated and accessible air transport system.

Connectivity, often measured using indices such as the Air Connectivity Index (ACI), reflects the strength and competitiveness of a country’s air transport network. Higher ACI scores indicate better integration into global networks, based on metrics such as the number of destinations served, service frequency, and available seat capacity \citep{Arvis2011}. However, many African countries lack the necessary data to produce a valid score, exposing persistent gaps in network accessibility across the continent. These gaps are attributed to factors such as high travel costs, poor safety records, corruption, infrastructure limitations, low internet penetration, and a shortage of skilled professionals \citep{Heinz2013}.

The International Air Transport Association (IATA) defines air connectivity as the ability to travel efficiently from origin to destination in the shortest possible time, emphasising that its measurement depends on the economic value generated by individual air linkages \citep{IATA2020}, i.e. airline supply decisions. Improved connectivity thus enhances access to goods, services, and people, which in turn fuels demand for air travel. 

Ethiopia for example demonstrates the benefits of strategic investment in air transport. Despite being landlocked, the country has emerged as the most internationally connected in sub-Saharan Africa. From 2007 to 2018, international trips grew by 49.4\% - compared to for example South Africa, which saw a 14.7\% decline over the same period \citep{Tolcha2020}. Ethiopia’s geographic location, coupled with the success of Ethiopian Airlines, has positioned it as a key hub linking sub-Saharan Africa to Europe, the Middle East, and Asia. Ethiopian Airlines has outperformed all other African state-owned carriers by maintaining a vast intra-African network, leveraging a strong hub for connecting traffic and forming strategic partnerships with regional airlines \citep{MEICHSNER2018}.

Many African state-owned carriers are struggling with profitability and operational sustainability. Airlines such as Air Afrique, Nigeria Airways, Air Gabon, and Ghana Airways Corporation collapsed in the early 2000s, resulting in a decline in intra-African connectivity despite overall capacity growth \citep{Bofinger2008}. The same study showed that between 2001 and 2007, 31 African airlines exited the market, removing 8 million seats, while 34 new entrants brought in nearly double that capacity. However, the number of origin-destination pair connections declined from 218 to 190, which shows a growing imbalance between capacity and actual connectivity. \cite{Gwilliam2011} notes that while air traffic volumes are rising, network development remains uneven and is often hindered by outdated airport infrastructure, unreliable state-owned carriers, and political barriers to market access. Most airports within Africa are not equipped to accommodate the growing demand for air travel or to serve as functional regional hubs. Addressing these deficiencies requires coordinated policy reform and investment in infrastructure.

Much of the lost connectivity can be attributed to the inability of national airlines to remain viable, along with weak infrastructure and slow and uneven liberalisation. Despite these setbacks, the liberalisation of southern African airspace through initiatives such as the Open Skies Agreement has started to bring benefits, including more airline options, increased service frequency and improved hub connections \citep{Mhlanga2017}. South African Airways is an example of a state-owned carrier that has persistently failed to implement a turnaround strategy, due to slow decision-making, undercapitalisation, and leadership instability \citep{Nyatsumba2023}. Cross-border air transport remains sensitive to political (in)stability, but liberalised markets have facilitated the emergence and expansion of new carriers.

\section{Literature}\label{literature}
This section discusses literature on airline capacity-supply decisions, the Yamoussoukro Decision aimed at liberalisation of air transport services among African countries, and the various factors influencing air connectivity across Africa. The section ends with a brief discussion of previous applications of choice modelling in aviation. 

While numerous factors potentially influence airline choices, four key determinants emerge consistently across aviation research: GDP and economic conditions, market size and urban populations, historical influences, and the impact of COVID-19 \citep{Suryani2010, Button2015, Boonekamp2018, Tirtha2023}, which we will discuss in detail.   

\subsection{Airline capacity-supply and route planning}
Route planning in the airline industry is a complex process aimed at maintaining profitability while ensuring connectivity is prioritised.  The structure of an airline's route network is a fundamental component of its business model \citep{Niehaus2009}.  According to \cite{Heinz2013}, connectivity is a key differentiator between airline business models as it reflects the network design, such as hub-and-spoke and point-to-point networks. Airlines must continuously evaluate the viability of routes, and  reductions in capacity-supply between specific origin-destination pairs often stem from declining profitability rather than demographic limitations. Fluctuations in travel demand — driven by economic shifts, tourism patterns, and competing routes offering lower fares or direct services — directly affect supply decisions. Profitability assessments must also account for market-related costs such as fuel prices, airport fees, labour, and maintenance, which vary across regions \citep{Zhou2020}.

Airlines optimise their seat capacity by adjusting flight frequency and aircraft size in response to (prognoses of) passenger demand. When demand is insufficient, flights may be reduced or cancelled to maintain high seat occupancy, although doing so may incur penalties for unused airport slots, an added complexity in strategic planning \citep{Yu2009}. Slot availability remains a key constraint in determining supply. Aircraft selection further influences supply and financial performance, with airlines often favouring smaller aircraft operated more frequently to maximise seat utilisation rates and service frequency \citep{Hansen2005}. However, this approach must be weighed against increased operating costs, including staffing and ground services. Additionally, profitability is strongly linked to the proportion of premium seats sold, with premium class passengers often being more desirable than larger volumes of economy-class travellers.

\subsection{Participation in the Yamoussoukro Decision}
Air transport in Africa has historically been constrained by fragmented infrastructure and restrictive bilateral air service agreements (BASAs), limiting intra-African connectivity and competition \citep{Njoya2016}. Following independence, many African states attempted to establish national carriers, but these efforts were often unsuccessful due to high operating costs, limited infrastructure, and market inefficiencies. Instead, intercontinental routes were prioritised, leaving the intra-African network underdeveloped. By 1990, only 57 bilateral agreements existed between African country pairs, compared to 249 with non-African states, highlighting a historical imbalance in air transport priorities \citep{Schlumberger2010}. 

Current challenges, including high user charges, limited infrastructure investment, and regulatory hurdles, persist despite increasing demand driven by urbanisation and economic development \citep{Njoroge2020}. These conditions highlight the urgent need for integrated and liberalised air transport strategies to enhance accessibility and drive economic growth across the continent.
The Yamoussoukro Declaration, adopted in July 2000, represents a critical step towards air transport liberalisation by replacing restrictive BASAs with a unified regulatory framework to enhance intra-African connectivity \citep{Njoya2016}. This granted third, fourth, and fifth freedom traffic rights and eliminated restrictions on airline ownership, tariffs, and scheduling, thereby allowing greater flexibility for African carriers \citep{AfricanUnion2020}: 
\begin{itemize}
     \item “Third freedom traffic right”: the right of an Eligible Airline of one State Party to put down, in the territory of another State Party, passengers, freight and mail taken up in the State Party in which it is licensed. 
    \item “Fourth freedom traffic right”: the right of an Eligible Airline of one State Party to take on, in the territory of another State Party, passengers, freight and mail for offloading in the State Party in which it is licensed. 
    \item “Fifth freedom traffic right”: the right of an Eligible Airline of one State Party to carry passengers, freight and mail between two State Parties other than the State Party in which it is licensed.
\end{itemize}
However, implementation has been slow and inconsistent, limited by political resistance, weak enforcement, and unresolved regulatory issues \citep{Schlumberger2010}. Recognising these barriers, African leaders have introduced the Single African Air Transport Market (SAATM)\footnote{The SAATM is an initiative aimed at advancing liberalisation and promoting connectivity within Africa, playing an important role in promoting economic, political and social integration in the continent \citep{IATA2020}} under the African Union's Agenda 2063 to revitalise liberalisation efforts. \cite{Abeyratne2003} notes the Yamoussoukro Decision's economic aspirations; particularly for post-conflict African states, seeking recovery and growth are significant. If fully realised, the Yamoussoukro Decision and SAATM could lower airfares, increase competition, enhance regional trade, and improve mobility. Nevertheless, its effectiveness is reliant on stronger regulatory enforcement and greater political commitment.

\subsection{Factors that can influence air connectivity}\label{factors}

\subsubsection{GDP and the economy}
The relationship between macroeconomic factors, particularly GDP, and airline supply is critical in understanding how the aviation industry adapts to changes in the economy. \cite{Ishutkina2009} analysed the interaction between air transport and economic activity. Their findings show that economic growth has caused people to shift their demand towards faster modes of transport, which also results in higher demand for air transport. These studies indicate that an increase in GDP is associated with an increase in both international and domestic airline supply. 

\subsubsection{Market size and urban populations}
In principle, countries with larger populations have greater transport demand, and depending on the country’s economy and air transport connectivity, air travel demand.  For instance, although Nigeria and Ethiopia are Africa’s most populous countries, South Africa and Kenya have significantly higher passenger-to-population ratios, indicating that economic development and urban concentration play a more decisive role in shaping air travel patterns \citep{Tolcha2020}. Countries with multiple urban hubs tend to have stronger domestic air networks, reinforcing the link between air transport demand and economic activity. \cite{Brueckner2003} further demonstrates that increased air services stimulate economic growth, showing that a 10\% rise in enplanements correlates with a 1\% increase in service-sector employment. Similarly, \cite{Marazzo2010} identified a long-term relationship between GDP and air travel demand in Brazil, confirming GDP and market size as key predictors of airline supply. \cite{Abate2018} argue that liberalisation fosters affordability and frequency, thereby boosting demand, although restrictive policies still hinder this potential across much of the continent. Strategic investment in infrastructure and supportive policy frameworks remain essential for aligning Africa’s air transport sector with its growing urban population.

\subsubsection{Historical influences on airline supply}
Historical and cultural ties play a critical role in shaping airline networks and travel patterns across Africa. Colonial legacies continue to influence air transport, as demonstrated by \cite{Button2015}, who found persistent connections between former British and French colonies in sub-Saharan Africa. These colonial-era linkages often translate into stronger aviation ties with historically affiliated countries, including more direct flights and higher service frequencies. Despite post-independence efforts to reform transport policies, the influence of colonial interests has endured. Many Sub-Saharan countries focused on developing international connections to and from their former colonisers \citep{Schlumberger2010}. Shared language groups further reinforce connectivity by fostering stronger social and cultural ties. Although less formally documented, these linguistic affinities possibly influence air travel and trade between countries with shared histories, making them a relevant consideration for airline network planning. Understanding these historical and cultural dynamics is vital to developing a more integrated and strategically connected African air transport system.

\subsubsection{Impact of COVID-19 and recovery}
The Coronavirus Disease 2019 (COVID-19) pandemic has undoubtedly affected the transport industry \citep{Amaris2024}, and aviation in particular. The pandemic caused major disruptions in both airline supply and travel demand. \cite{Mazareanu2021} reported a 50\% decline in scheduled passengers in 2021 compared to 2019, reflecting the severity of the disruption. Similar to earlier shocks such as 9/11 and the SARS outbreak, the pandemic's effects extended beyond the short term, as shown by \cite{Chi2013}, who found that air passenger demand is susceptible to economic and security-related crises. \cite{Suau-Sanchez2020} and \cite{Zhang2024} note that partial travel bans, including border closures and quarantine measures, significantly delayed the recovery process, making it more severe than previous health emergencies. According to \cite{ICAO2023}, airlines reduced their seat offerings by 47–58\% in the first half of 2020, a clear indication of the scale of operational cutbacks. These findings highlight the vulnerability of the air transport sector to global health crises and the importance of resilience in airline supply strategies.

\subsection{Application of choice modelling in aviation}

The early study by \cite{Akiva1987} laid the foundation for numerous choice modelling studies in transport, including in aviation. Some important aviation choice modelling studies have since been published by \citet{Proussaloglou1999} on flight choice, by \citet{Hensher2001} on airline choice, and by \citet{Hess2005} on airport choice, among others. 

\cite{Proussaloglou1999} developed air traveller discrete choice models to better understand the trade-offs passengers make when selecting carriers, flights and fare classes – all of which can inform airline decisions on scheduling, pricing and seat allocation. Their framework examines passenger demand in a competitive market, quantifying the impact of carrier preference, carrier loyalty, departure and arrival time preferences, and willingness-to-pay (WTP) to avoid delays. The findings highlight the price premiums that business and leisure travellers are willing to pay to avoid schedule delays, fly with their preferred airline, and have fewer travel restrictions. 

Airline choices are investigated by \cite{Hensher2001} using a complex stated choice experiment with 32 choice tasks, testing variations by presenting different profiles to participants deciding whether to fly between Australia and New Zealand. Their analysis explores new opportunities for patronage in air travel by using a strong combination of stated choice methods with revealed preference models. 

A study by \cite{Hess2005} analysed airport choices based on factors such as access time, travel purposes, and random variations within groups of travellers.  Their mixed logit study revealed that business travellers are less sensitive to fare increases than leisure travellers and are willing to pay more for quicker access, highlighting differences in sensitivity to key travel attributes. Building on this, \cite{Kolker2024} contribute to understanding global passenger flows by developing discrete choice models that analyse airline passenger decisions based on factors such as ticket price, travel time and transfer frequency. Their study used passenger demand data to identify elasticities and passenger flows, emphasising how passenger decisions drive airline operations and shape global air transportation systems. 

As evidenced by the literature review, air travel modelling has traditionally focused on passenger demand, analysing travellers' preferences, willingness to pay, and sensitivity to fares. While these studies have advanced our understanding of travel behaviour, they do not give insight into supply decision-making processes that ultimately shape air connectivity. 

Building on the foundations laid by previous studies, this paper contributes to the literature by shifting the perspective to the airlines' supply decisions, specifically in the context of African-based carriers. This approach provides valuable insights into how economic, geographic and cultural factors influence the distribution of air services. 
By focusing on supply behaviour, this research aims to inform policy frameworks and strategic interventions that support a more integrated and competitive African air transport market. 

\section{Data}\label{data}

To investigate the factors that influence air connectivity in Africa, OAG data has been utilised in conjunction with relevant secondary data.

\subsection{Official Aviation Guide data}

The large-scale data provided by the Official Aviation Guide (OAG) consists of a flight database that includes all regular passenger direct flights operated by airlines worldwide during the first full week of November, spanning five years (2016, 2018, 2020, 2021 and 2022), which includes datapoints from before, during and immediately after the COVID-19 pandemic. 

The data for the present paper was extracted from PTV VISUM in a format that only considered African countries in the origin-destination (OD) pairs. An O-D pair refers to a link that connects two locations, namely the start and end points of a journey. The Africa-focused database comprises 267 international airports. The thickness of the red line in Figure \ref{fig:africa} indicates the relative seat capacity available between each pair of cities with an international airport. 

The network map illustrates the continental connectivity across Africa, displaying 170 country pairs and revealing a hub-and-spoke structure with several dominant hubs. 
\begin{itemize}[left=1.5em,labelsep=0.5em,itemsep=-2pt]
    \item The North African hubs, particularly Egypt and Morocco, shows substantial capacity and connections to other African cities, reflecting economic importance.
    \item The West African hubs, connecting Nigeria, Ghana and Ivory Coast, displays a robust network and serves as a connection to landlocked nations.
    \item The East African hubs, centred on Ethiopia and extending through Kenya, demonstrate significant capacity and serve as the primary gateway for East Africa. This also connects the region to western, northern, and southern hubs.
    \item The Southern African hubs, radiating mostly from South Africa, show extensive connectivity throughout the region, with particular strong links towards East and West African hubs. 
\end{itemize}

\subsection{Data adaptation for present paper}

The primary database includes codeshare flights. Codeshare flights have become a competitive feature in the industry, offering more flight options. This availability generally helps passengers by simplifying the booking process and enhancing the feasibility of aviation management \citep{Gayle2007}.  There are non-African airlines that operate among African countries; however, it is important to note that these trips may be part of a greater journey and operated by a different partner airline. The database excluded duplicate information from codeshare flights to ensure the validity of seat capacity figures. 

\begin{figure}
    \centering
    \includegraphics[width=1\linewidth]{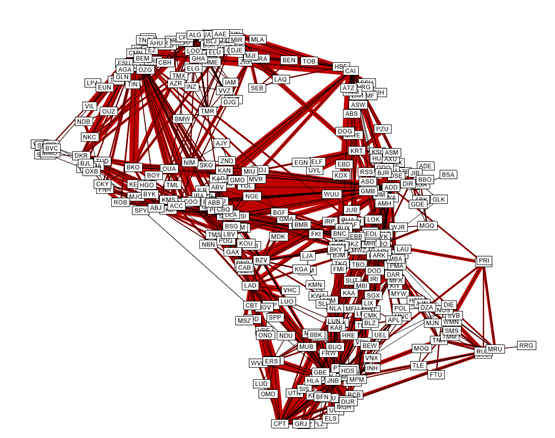}
    \caption{Capacity-supply in terms of seats offered between O-D pairs within Africa during one week in November 2016.}
    \label{fig:africa}
\end{figure}

For the purpose of the present paper, we also performed an additional level of aggregation. While Figure \ref{fig:africa} shows all city pairs, the database was simplified to work with country pairs instead of city pairs. 

The focus of the present paper is on understanding connectivity, and in particular what determines capacity on given routes. With this in mind, the database is structured with origin-destination (OD) country pairs as the observations, and with the airlines being the alternatives. 

Using this structure, the database enables a primary analysis of seat availability and flight frequency trends for intra-African international travel relative to non-African carriers. The data in Table \ref{table:seats_freq} highlights the market share of African carriers compared to European, Asian and Middle Eastern competitors in the African context. While African carriers maintained a dominant share of seat capacity throughout the period, their position weakened from 2016 to 2022. This contraction, with the sharpest drop occurring in 2020 during the COVID-19 pandemic, contrasts with non-African carriers' resilience, which actually expanded capacity during the pandemic period. This suggests that non-African carriers capitalised on the reduced presence of African airlines during the pandemic and seized the opportunity to fill market gaps.

The flight frequency data reveals a distinct recovery pattern. By 2022, African carriers had restored flight frequencies to near pre-COVID-19 levels, while seat capacity for African carriers remained at only 46\% of 2016 levels. This gap indicates a fundamental shift toward smaller aircraft operations, suggesting African airlines are managing capacity constraints through smaller equipment rather than maintaining previous levels of seat capacity per route. 

\begin{table}[h]
    \centering
    \caption{Weekly seat availability and flight frequencies for intra-African travel}
    \label{table:seats_freq}\resizebox{\linewidth}{!}{
    \begin{tabular}{lccccc}
         \hline
         \textbf{Metric} & \textbf{2016} & \textbf{2018} & \textbf{2020} & \textbf{2021} & \textbf{2022}\\
         \hline
         \multicolumn{6}{l}{\textit{Seat Availability}} \\
         African carriers & 301,815 & 220,730 & 115,189 & 140,330 & 138,330\\
         Non-African carriers & 30,882 & 34,048 & 37,286 & 31,362 & 34,200\\
         \% African carriers & 90.7\% & 86.6\% & 75.5\% & 81.7\% & 80.2\% \\
         \hline
         \multicolumn{6}{l}{\textit{Flight Frequencies}} \\
         African carriers & 4,075 & 4,392 & 2,866 & 2,903 & 3,684\\
         Non-African carriers & 276 & 308 & 471 & 223 & 232\\
         \% African carriers & 93.7\% & 93.4\% & 85.9\% & 92.9\% & 94.1\% \\
         \hline
    \end{tabular}}
\end{table}

\subsection{Additional data}

For the purposes of our analysis, the attributes related to each O-D pair need to capture the key factors influencing connectivity that this study explores. We thus added GDP and urban population figures into the database, sourced from the World Bank open data catalogue.

In addition, we added details on colonial ties, shared language groups, distance, and indicators for neighbouring country relationships, all of which are hypothesised to shape supply decisions, as discussed in section \ref{factors}. Finally, for each O-D airline triplet, we also added a variable to indicate whether either the origin or destination was the home country for that airline.

\section{Modelling methodology}\label{method}

We use choice models to analyse the data described above. The work done by \cite{DanielMcfadden1973} set the foundation for discrete choice modelling. A distinctive feature of these models is their consideration of a rich pattern of influences on behaviour. Preferences are shaped by multiple factors, including characteristics of the decision maker, their attitudes and perceptions towards the alternatives at hand, as well as the actual attributes of the alternatives and how decision makers value those. These models are invaluable not only for predicting transport choices but also for informing policy and improving transportation planning by aligning systems with the needs and preferences of its users.

The insights gained will provide a comprehensive representation of how airlines determine their presence and capacity allocation on intra-African routes, offering insights into the dynamics of airline-driven connectivity within the continent.

Our specific focus is on the capacity different airlines offer on different O-Ds. For a given O-D, each airline decides on whether or not to operate, and how many seats to make available. The modelling problem in this study can thus be characterised as a discrete-continuous choice problem, which we will discuss next. 

\subsection{Discrete-continuous models}

Discrete-continuous choice models deal with choice situations where decision-makers face multiple alternatives and can select more than one option at the same time and consumer more than one unit per option \citep{Hanemann1984}, i.e. in cases where choices are not just limited to selecting one alternative (`whether to choose'), but also determining the extent of its use (`how much to consume'). As such, these models capture the interdependence between discrete and continuous decisions, which allows a more holistic understanding of behaviour. 

Recent discrete-continuous choice models include the Multiple Discrete-Continuous Extreme Value (MDCEV) model \citep{Bhat2005} and the extended Multiple Discrete Continuous (eMDC) model \citep{Palma2022}, amongst others. 

The MDCEV model \citep{Bhat2005} serves as a stochastic representation of the consumer maximisation process, wherein consumers allocate their resources in a manner that maximises their utility. This is a very flexible framework which remains grounded in micro-economic theory and has direct links to the Generalised Extreme Value (GEV) family of discrete choice models \citep[cf.][]{Train_2009}. However, the MDCEV framework has a key feature that presents challenges in the current application. Specifically, it relies on a budget constraint, which refers to the limitation on the resources available to the decision-maker(s) when choosing alternatives. This type of constraint is natural in the context of consumer decisions but presents difficulties when modelling allocation of seats in an air travel context. In the current study, a proxy for the budget constraint could be interpreted as relating to the total number of slots available in the origin or destination airport. However, we have no access to such data, and the number of slots per se is not equivalent to the number of possible seats given differences in aircraft size.

As such, directly applying MDCEV would impose restrictive assumptions that may not reflect actual airline supply behaviour. The eMDC model therefore provides a more flexible and realistic framework for capturing airline supply decisions, enabling the analysis of how the seat capacity for a given O-D pair is distributed across airlines without imposing constraints on total capacity.

The extended Multiple Discrete Continuous (eMDC) model differs from the MDCEV by incorporating the concepts of complementarity and substitution and by not requiring a budget constraint. Complementarity and substitution describe the interrelationships between the demand for pairs of products. Specifically, complementarity occurs when an increase in the demand for one good leads to an increase in the demand for another good. In contrast, substitution occurs when an increase in the demand for one good leads to a decrease in the demand for another. These concepts, originally articulated by \cite{Hicks1934}, are essential for achieving a more precise representation of consumer behaviour \citep{Palma2022}. 

Let us assume an outside good $0$ with $K$ inside goods ($k=1,\hdots,K$). An individual $n$ then decides what products $k$ to consume from a set of alternatives, by maximising his or her utility subject to a budget constraint:
	
	\begin{eqnarray}\label{eqOptimProb}
	\text{MAX}_{x_n} & & u_{0}(x_{n0}) + \sum_{k=1}^{K}u_{k}(x_{nk}) + \sum_{k=1}^{K-1}\sum_{l=k+1}^{K}u_{kl}(x_{nk},x_{nl})\\
	\nonumber \text{s.t.}      & & x_{n0}p_{n0} + \sum_{k=1}^{K}x_{nk}p_{nk} = B_n,
	\end{eqnarray}

where $x_n=[x_{n0}, x_{n1}, ..., x_{nK}]$ is a vector grouping the consumed amount of each product, $p_{nk}$ is the price of product $k$ for individual $n$, and $B_n$ is the total budget available to $n$. The consumption $x_{n0}$ is an \textit{outside} or \textit{numeraire} good, i.e. a good that aggregates all consumption outside of the category of interest. In our case, this would represent all unallocated capacity. 

The components of the utility equation for the eMDC assume the following functional forms:

\begin{align}
u_0(x_{n0}) &= \psi_{n0} x_{n0}\\
u_k(x_{nk}) &= \psi_{nk} \gamma_k \log\left(\frac{x_{nk}}{\gamma_k} + 1\right)\\
u_{kl}(x_{nk}, x_{nl}) &= \delta_{kl}(1 - e^{-x_{nk}})(1 - e^{-x_{nl}})
\end{align}

where $\psi_{n0} = e^{\alpha z_{n0}}$ and 
$\psi_{nk} = e^{\beta_k z_{nk} + \varepsilon_{nk}}$ are the baseline utilities (i.e. marginal utility at zero consumption) of the different goods. We have that:

\begin{itemize}[left=1.5em,labelsep=0.5em,itemsep=-2pt]
    \item $z_{n0}$ are characteristics of person $n$
    \item $z_{nk}$ are characteristics of person $n$ and product $k$.
    \item $\alpha$ captures the effect of explanatory variables $z_{n0}$ on the marginal utility of the outside good, where this can also include product-specific constants.
    \item $\beta_k$ captures the effect of explanatory variables $z_{nk}$ on the marginal utility of good $k$.
    \item $\varepsilon_{nk}$ is a random disturbance following a Normal distribution with a mean of $0$ and a common standard deviation $\sigma$.
    \item  the $\gamma_{k}$ parameters relate mainly to consumption satiation, by altering the curvature of alternative $k$'s utility function. In general, a higher $\gamma_k$ indicates higher consumption of alternative $k$, when consumed.
    \item $\delta_{kl}$ parameters measure complementarity between alternatives $k$ and $l$ if $\delta_{kl} > 0$, and substitution if $\delta_{kl} < 0$.
\end{itemize}

The assumption of a linear utility for the outside good allows us to drop the outside good and the budget assumption \citep[cf.][]{Palma2022}. The resulting likelihood function $L$ for the eMDC is then:

\begin{equation}
L(x_k) = |J| \prod_{k=1}^K f(-W_k)^{I_{x_k>0}} F(-W_k)^{I_{x_k=0}}
\end{equation}
where
\begin{equation}
\begin{aligned}
W_k &= \beta_k z_{nk} - \ln\left(\frac{x_{nk}}{\gamma_k} + 1\right) \\
&\quad - \ln\left(\psi_{n0} p_{nk} - e^{-x_{nk}} \sum_{l \neq k} \delta_{kl}\left(1 - e^{-x_{nl}}\right)\right)
\end{aligned}
\end{equation}

The Jacobian elements are:
\begin{equation}
J_{ii} = \frac{1}{x_i + \gamma_i} + \frac{E_i}{\psi_0 p_i - E_i}
\end{equation}

\begin{equation}
J_{ij} = \frac{-\delta_{ij}e^{-x_i} e^{-x_j}}{\psi_0 p_i - E_i}
\end{equation}

where
\begin{equation}
E_i = e^{-x_i} \sum_{l \neq i} \delta_{il}(1 - e^{-x_l})
\end{equation}

\begin{itemize}
    \item  $|J|$ is the value of the determinant of the Jacobian $J$ of vector $-W_{k}$ where $k$ implies consumed alternatives. 
    \item $I_{x_{k}>0}$ and $I_{x_{k}=0}$ are binary variables taking a value of 1 if $x_{k}>0$ or $x_{k}=0$, respectively, or zero otherwise. 
\end{itemize}

\subsection{Implementation for our case study}

To estimate capacity-supply decisions across airlines using the eMDC model, we rely on Apollo \citep{Hess2019}, using the Bunch-Gay-Welsch (BGW) algorithm for estimation \cite{Bunch1993}. 
The observation $n$ is the origin-destination pair, and the dependent variable is the airline seat capacity $x_k$, with airlines $K$ as the alternatives. 

Two eMDC models were estimated to analyse various aspects of airline attractiveness and supply in the African aviation market. The models did not incorporate complementarity and substitution. This simplification is largely due to our work looking at O-D pairs as observations and airlines as alternatives, where our assumption in essence implies that capacity allocation decisions for a given route are made independently across airlines. 

Model 1 is our base eMDC model, which uses covariates only for the airlines, in particular grouping them geographically by region. This model helps us understand how the capacity decisions made by African-based airlines in the broader African aviation market differ from those made by airlines from other continents. These terms were used both in the baseline utilities ($\beta$ parameters) and in the satiation component ($\gamma$ parameters).

The satiation parameters were specified as in Equation \ref{gamma}. Each airline is assigned a regional base satiation parameter ($\gamma_{Africa}$, $\gamma_{Asia}$, $\gamma_{Middle East}$ or $\gamma_{Europe}$) depending on its base country. This is then multiplied by a number of potential interactions with covariates, where these interactions are used inside an exponential transform to ensure that $\gamma_{nk}>0$.

\begin{equation}\label{gamma}
\gamma_{nk} = \gamma_k \cdot \exp\left(\zeta'\cdot z_{nk}\right)
\end{equation}

where $\zeta$ are the parameters capturing the impact of covariates on $\gamma$ in Model 2. 

Model 2 extends on this by in addition incorporating O-D pair and timepoint characteristics, thus helping us understand whether specific characteristics make it more or less likely that capacity is offered. We include a number of characteristics in this, as follows:
\begin{description}
    \item[GDP per capita:] this is used both in terms of the sum between the two countries (origin and destination) as well as the product, thus also capturing differences within a pair;
    \item[Population:] this is again used both in terms of the sum between the two countries as well as the product;
    \item[Urban population:] this is the proportion of the total population living in urban regions; 
    \item[Home country:] this is a dummy term included when either the origin or the destination is the home country of the airline. This term is used to understand the implementation and effectiveness of the Fifth Freedom Traffic Right in the Yamoussoukro Decision;
    \item[Neighbouring:] this is a dummy term included if the origin and destination are neighbouring countries;
    \item[Distance:] this is the distance between the capitals of the two countries;
    \item[Shared colony:] this is a dummy term if the two countries have a shared colonial history;
    \item[Shared language:] this is a dummy term if the two countries share a common language;
    \item[Time point:] A dummy term was included for post-COVID-19 years.
\end{description}

Again, all these terms were used both in the $\beta$ and $\gamma$ parts of the model. With the exception of the \textit{home country} term which is used in the $\beta$ component only.

\section{Results and Interpretation}\label{results}


\begin{table*}[t]
  \centering
  \caption{eMDC Choice Model Results}\label{tab:results_table}
    \begin{tabular}{llrrrr}
          &       & \multicolumn{2}{l}{\textbf{Model 1}} & \multicolumn{2}{l}{\textbf{Model 2}} \\
    \hline
    $LL$ at convergence  &       & -2473 &       & -2074 &  \\
    \hline
          & \textbf{Parameters} & \multicolumn{1}{c}{\textbf{Estimate}} & \multicolumn{1}{c}{\textbf{Robust t-ratio}} & \multicolumn{1}{c}{\textbf{Estimate}} & \multicolumn{1}{c}{\textbf{Robust t-ratio}} \\
    \multicolumn{1}{l}{\multirow{4}[2]{*}{\makecell[l]{$\beta$ parameters:\\ Airlines grouped by region}}}
        & $\beta_{Africa}$ & 0.061 & 2.249 & 0.000 & \multicolumn{1}{r}{NA} \\
          & $\beta_{Asia}$ & -0.251 & -3.162 & -0.073 & \multicolumn{1}{r}{-1.258} \\
          & $\beta_{Europe}$ & -0.314 & -3.799 & -0.053 & \multicolumn{1}{r}{-0.785} \\
          & $\beta_{Middle East}$ & -0.552 & -6.106 & -0.295 & \multicolumn{1}{r}{-4.742} \\
    \hline
    \multicolumn{1}{l}{\multirow{10}[1]{*}{\makecell[l]{$\beta$ parameters: \\ impact of O-D pair characteristics}}} 
        & $\beta_{GDP\_per\_capita\_sum}$ &       &       & -0.362 & \multicolumn{1}{r}{-2.139} \\
        &     $\beta_{GDP\_per\_capita\_prod}$ &       &       & 0.889 & \multicolumn{1}{r}{1.896} \\
          &     $\beta_{population\_sum}$ &       &       & -0.264 & \multicolumn{1}{r}{-5.094} \\
          &     $\beta_{population\_prod}$ &       &       & 0.190 & \multicolumn{1}{r}{1.638} \\
          &     $\beta_{urbanpopulation}$ &       &       & 0.172 & \multicolumn{1}{r}{0.796} \\
          &     $\beta_{homecountry}$ &       &       & 0.637 & \multicolumn{1}{r}{10.965} \\
          &     $\beta_{neighbouring}$ &       &       & 0.055 & \multicolumn{1}{r}{1.298} \\
          &     $\beta_{distance}$ &       &       & 0.017 & \multicolumn{1}{r}{2.427} \\
          &     $\beta_{sharedcolony}$ &       &       & -0.091 & \multicolumn{1}{r}{-2.163} \\
          &     $\beta_{sharedlanguage}$ &       &       & -0.078 & \multicolumn{1}{r}{-1.660} \\
    \multirow{2}[1]{*}{Time point factor} 
        & $\beta_{preCovid}$ &       &       & 0.000 & \multicolumn{1}{r}{NA}\\
          & $\beta_{postCovid}$ &       &       & 0.346 & \multicolumn{1}{r}{5.716} \\
    \hline
    \multicolumn{1}{l}{\multirow{4}[2]{*}{\makecell[l]{$\gamma$ parameters: \\ Airlines grouped by region}}} 
        & $\gamma_{Africa}$ & 1.167 & 8.662 & 0.403 & \multicolumn{1}{r}{6.640} \\
          & $\gamma_{Asia}$ & 1.624 & 5.653 & 0.860 & \multicolumn{1}{r}{4.645} \\
          & $\gamma_{Europe}$ & 1.242 & 5.636 & 0.653 & \multicolumn{1}{r}{4.828} \\
          & $\gamma_{Middle East}$ & 4.102 & 5.644 & 2.071 & \multicolumn{1}{r}{4.842} \\
    \hline
    \multicolumn{1}{l}{\multirow{10}[1]{*}{\makecell[l]{$\zeta$ parameters: \\ impact of O-D pair characteristics on $\gamma$}}} 
        &     $\zeta_{GDP\_per\_capita\_sum}$ &       &       & 0.343  & \multicolumn{1}{r}{1.136} \\
          &     $\zeta_{GDP\_per\_capita\_prod}$ &       &       & -0.004 & \multicolumn{1}{r}{-0.455} \\
          &     $\zeta_{population\_sum}$ &       &       & 0.046 & \multicolumn{1}{r}{4.344} \\
          &     $\zeta_{population\_prod}$ &       &       & 0.220 & \multicolumn{1}{r}{0.086} \\
          &     $\zeta_{urbanpopulation}$ &       &       & 0.509 & \multicolumn{1}{r}{1.135} \\
          &     $\zeta_{neighbouring}$ &       &       & 0.181 & \multicolumn{1}{r}{2.102} \\
          &     $\zeta_{distance}$ &       &       & 0.045 &  \multicolumn{1}{r}{2.749} \\
          &     $\zeta_{sharedcolony}$ &       &       & 0.205 &  \multicolumn{1}{r}{2.355} \\
          &     $\zeta_{sharedlanguage}$ &       &       & 0.290 & \multicolumn{1}{r}{2.766} \\
    \multirow{2}[0]{*}{Time point factor} 
        &     $\zeta_{preCovid}$ &       &       & 0.000 & \multicolumn{1}{r}{NA}\\
          &     $\zeta_{postCovid}$ &       &       & -0.676 & \multicolumn{1}{r}{-7.155} \\
    \hline
    Scaling parameter    & {$\sigma$} & 0.642 & 12.012    & 0.497 & \multicolumn{1}{r}{13.149}\\
    \end{tabular}%
\end{table*}%


The model results are summarised in Table \ref{tab:results_table}. The log-likelihood (LL) values provide an indication of the model fit to the observed data. A higher log-likelihood value indicates better model fit, suggesting that the model captures the underlying behaviour of airline supply more accurately. Model 2 shows a significant improvement of 399 units in model fit over Model 1, with a Likelihood Ratio (LR) test value of 798 which gives us $p< 0.001$, thus clearly rejecting the null hypothesis that the two models are equal. Model 2 includes an additional 12 $\beta$ covariate parameters and 11 $\zeta$ covariate parameters. This shows that incorporating economic, geographic and cultural covariates enhances the explanatory power of the models, providing a better understanding of capacity decisions. 

We interpret this model by comparing the estimates within each parameter group to indicate their relative importance. The interpretation is mainly focused on Model 2. The results show that African-based airlines consistently have higher beta ($\beta$) values, indicating a greater intrinsic likelihood of offering capacity. This suggests that, before accounting for other variables, African carriers are inherently more likely to serve a greater number of routes within the continent. This reflects both their regional familiarity and operational base, making them more likely to allocate capacity to these routes, even in the absence of external influences.

African-based airlines have a lower gamma value ($\gamma$ = 0.403, t-ratio = 6.640), meaning that while they may be more likely to offer capacity (i.e. $x_k>0$) on any given route (as seen in the $\beta$ parameter), the level of satiation is higher, i.e. they are likely to offer lower capacity than is the case for other carriers. Notably, airlines from the Middle East exhibit higher gamma values ($\gamma$ = 2.071, t-ratio = 4.842), implying a slower rate of diminishing utility. This reflects their ability to supply large capacities between African country pairs, possibly due to their larger fleet sizes, extensive route networks and perceived service quality. 

One of the most significant findings is the positive effect of an airline’s home country ($\beta$ = 0.637, t-ratio = 10.965) on its utility, indicating that airlines are more likely to offer service if the origin or destination is their home country. Other than logistical advantages, this could imply significant preferential treatment for national carriers. This highlights the limited implementation and effectiveness of the Fifth Freedom Traffic Right in the Yamoussoukro Decision, which was intended to promote cross-border service and greater regional integration within the continent. 

\begin{figure}
    \centering
    \includegraphics[width=1\linewidth]{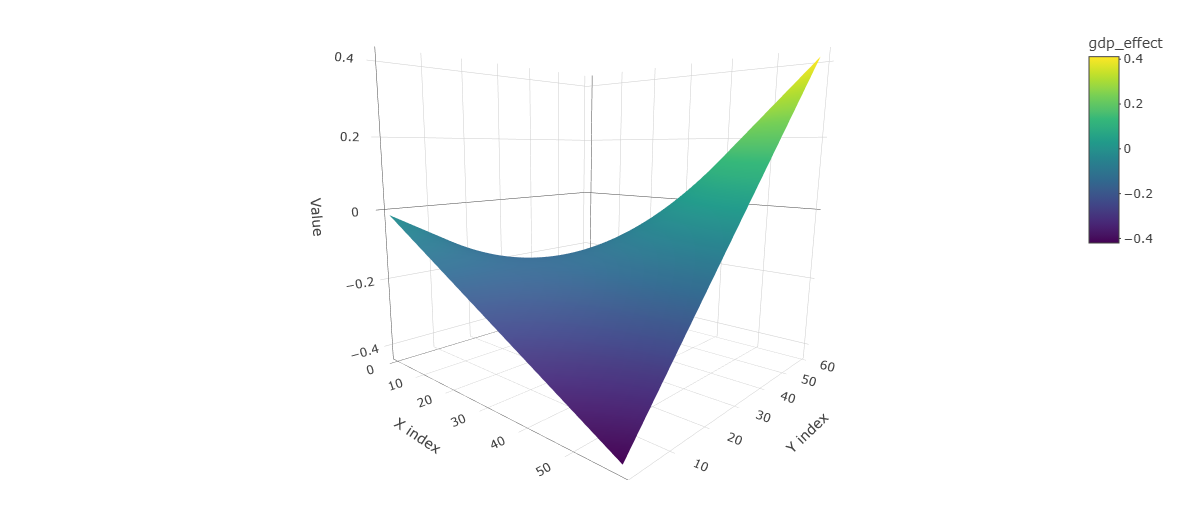}
    \caption{GDP per capita $\beta$ effect on air transport supply within Africa}
    \label{fig:gdp}
\end{figure}

Among the covariates included in the eMDC model, the GDP per capita product effect ($\beta$ = 0.889, t-ratio = 1.896) emerges as the most influential. This suggests that GDP per capita has a positive effect on whether seats are made available. It also implies that airlines increase supply when both countries are moderately wealthy, indicating profitable middle-market opportunities.  Figure \ref{fig:gdp} visually demonstrates this effect for the $\beta$ component. The x-axis plots the scaled GDP per capita value for the origin, and the y-axis plots the scaled GDP per capita value for the destination. The three-dimensional surface shows how the pairing of two economies’ GDP per capita influences the effect value.

In country pairs with larger combined populations ($\zeta$ = 0.046, t-ratio = 4.344), airlines tend to allocate more seats per route. This reflects supply-side responsiveness to market size, since larger populations often justify greater investment in capacity and route frequency. Airlines are more likely to deploy higher-capacity aircraft or increase service frequency in these markets, leveraging existing infrastructure to meet demand efficiently.

The post-COVID coefficient ($\beta$ = 0.346, t-ratio = 5.716) suggests airlines increased the number of routes they serve post-pandemic, indicating a recovery in supply. However, the negative zeta coefficient ($\zeta$ = -0.676, t-ratio = -7.155) suggests this recovery implied lower per route seat numbers, possibly due to fleet limitations. This finding aligns with the preliminary analysis in Table \ref{table:seats_freq}.

Routes between country pairs with shared colonial history receive more supply allocation ($\zeta$ = 0.205, t-ratio = 2.355). This indicates that the links established by past colonial powers continue to influence contemporary transport networks. Similarly, shared linguistic histories also receive more supply allocation ($\zeta$ = 0.290, t-ratio = 2.766) likely by reducing communication barriers. These findings demonstrate that cultural familiarity, through colonial ties and common language, is a crucial factor in shaping airline supply decisions.

\section{Conclusion, discussion and limitations}\label{conclusion}

This study provides an analysis of the factors that influence airline supply within the African aviation market. The significance of the results across the two extended Multiple Discrete Continuous (eMDC) models adds to the validity of the study. The results indicate that African-based airlines dominate intra-African routes in terms of capacity-supply across all models when compared to European, Asian and Middle Eastern airlines. 

The literature explored several key factors that affect air connectivity, including economic, geographic, cultural and external shocks. GDP is consistently shown to be a strong driver of air connectivity. Market size and GDP significantly shape passenger demand and airline supply.  Geographically strategic airports have the proximity advantage of forming a hub, though it is not the strongest influence on air connectivity. Cultural ties, especially former colonial relationships, also enhance direct flights and aviation links. Additionally, external shocks such as the COVID-19 pandemic have led to reductions in capacity offered by airlines.  

Based on the modelling results, it can be concluded that GDP per capita has a strong influence on airline utility and supply across Africa.  The GDP result highlights further complex market dynamics. African supply decisions appear to target pairs that either combine two high-income countries, or a middle-income and high-income country, while avoiding combinations of low and high-income countries. This reflects a strategic balance between affordability, competition, and operating costs in a market where growth is unevenly distributed \citep{Bowen2019}. The strong preference for middle to high income pairs suggests airlines target emerging corridors where competition is manageable. Underdeveloped countries in Africa may continue to be underserved, falling outside of the supply decisions, which could perpetuate isolation and unequal connectivity. 

The strong home country coefficient reinforces findings by \citep{Schlumberger2010} who note that the implementation of the Yamoussoukro Decision has been slow due to political resistance and regulatory issues. The dominance of national carriers can explain the lack of effectiveness of the fifth freedom traffic rights. This is also evident by the reduced attractiveness of foreign carriers relative to African carriers. This finding provides evidence that Africa's attempt at aviation liberalisation has had shortcomings in achieving its connectivity goals, despite efforts by SAATM. Implications can be extended to understanding Bilateral Air Service Agreements (BASAs), which continue to favour national carriers, causing a web of restricted routes. The relatively low beta coefficient ($\beta$ = -0.295, t-ratio = -4.742) for Middle Eastern carriers suggests that even efficient, well-capitalised airlines face systemic disadvantages in the intra-African air transport market.  

Cultural ties, including shared language groups, enhance capacity allocation on routes. These ties foster smoother communication, trust, and cooperation among stakeholders. As a result, cultural affinities represent an important factor in supply allocation decisions within Africa.

The post-COVID negative zeta effect indicates that the number of seats on routes decreased, which aligns with the global trend of capacity reductions \citep{Mazareanu2021, Tolcha2023}. The positive post-COVID-19 beta effect suggests that airlines adopted more strategic responses to market recovery, reflecting coordinated policy responses across African states.  

The findings from the study can inform policy recommendations and strategic decisions for airlines and stakeholders involved in the aviation industry to improve connectivity to and from African countries. These efforts can foster a more integrated and economically vibrant Africa through better alignment of airline services with market demands. 

\subsection*{Limitations}
While this paper provides valuable insights into intra-African airline decisions, it excludes domestic flights, flights to and from other continents, and popular indirect connections. As a result, the findings do not reflect the full spectrum of air connectivity, particularly the impact of central hubs and global connectivity on African routes.

Another key limitation is the exclusion of code-share flights. To maintain the integrity of the database and avoid duplicate capacity-supply entries, the study does not account for airline alliances. Consequently, the potential influence of partnerships and shared networks on supply decisions for route profitability and airline competition is not analysed.

The data supplied by OAG (Official Aviation Guide) contains all regular passenger flights of airlines worldwide for the first full week in November. The November-only data could invite seasonal bias and does not offer a complete representation of airline operations. The data could also underestimate the role of key hubs within Africa.  

Additionally, cargo supply is not considered in the analysis due to the uncertainty of the space it occupies within commercial flights. While passenger aircraft often transport both passengers and cargo, the study focuses solely on passenger capacity, which may not fully capture the economic viability of certain routes. 

This study did not look at competition from long-distance rail and bus travel within the continent, both of which serve as important competitors to air transport. Analysing these alternative modes could provide a more comprehensive understanding of general transport connectivity across Africa. This presents an opportunity for future research to assess the extent of competition between air travel and ground-based transport systems.

Obviously, this study looked at capacity-supply decisions, not at passenger decisions. However, there is a strong correlation between demand and what capacity is offered in an air travel context. Nevertheless, future research incorporating passenger decisions based on revealed or stated preference surveys could enhance understanding of the factors that influence air travel choices and ultimately capacity-supply decisions in Africa.

Despite these limitations, the study offers a strong foundation for understanding capacity supply decisions within the African context. Addressing these gaps in future research could further refine air transport modelling and policy recommendations.

\section*{Acknowledgements}
 Stephane Hess acknowledges the support of the European Research Council through advanced Grant 101020940-SYNERGY.

\section*{Declaration of generative AI and AI-assisted technologies in the writing process}
During the preparation of this work the primary author used Grammarly Pro in order to fix sentence structure and grammatical errors. After using this service, the authors reviewed and edited the content as needed and takes full responsibility for the content of the published article.

\appendix


\bibliographystyle{elsarticle-harv} 
\bibliography{example}



\end{document}